\documentclass[aps,pra,showpacs,fleqn,twocolumn]{revtex4}
\usepackage{graphicx}

\usepackage{amsmath}
\usepackage{amsfonts}
\usepackage{amssymb}

\usepackage[svgnames]{xcolor}

\newcommand{\Tr}{\text{Tr}\,}

\newcommand{\RDM}{\rho}
\newcommand{\ket}[1]{\left|#1\right\rangle}
\newcommand{\bra}[1]{\left\langle#1\right|}

\newcommand{\norm}[1]{\left\|#1\right\|_1}

\renewcommand{\Re}{\text{Re}\,}
\renewcommand{\imath}{\mathrm{i}}

\begin{document}
	\title{Nonlinear-dissipation-induced Entanglement of Coupled
          Nonlinear Oscillators}
		 
  \author{Aurora Voje}\email{aurora@chalmers.se}
  \author{Andreas Isacsson} 
  \author{Alexander Croy} \email{croy@chalmers.se} 
  \affiliation{Department of Applied Physics, Chalmers University of Technology, 
  					SE-412 96 G\"{o}teborg, Sweden}

\begin{abstract}
The quantum dynamics of two weakly coupled nonlinear oscillators is
analytically and numerically investigated in the context of nonlinear
dissipation. The latter facilitates the creation and preservation of
non-classical steady states. Starting from a microscopic description
of two oscillators individually interacting with their dissipative
environments, it is found that in addition to energy relaxation,
dephasing arises due to the mutual coupling. Using the negativity as
an entanglement measure, it is shown that the coupling entangles the
oscillators in the long-time limit. For finite temperatures,
entanglement sudden death and rebirth are observed.
\end{abstract}

\maketitle

\section{Introduction}
The counter-intuitive concept of dissipation-induced generation of
quantum states has recently attracted a lot of interest. It has been
proposed to create pure states \cite{dimi+08} and, in particular,
entangled states \cite{krbu+08,vewo+09,kare+11,krmu+11,kowi+12}.

Entanglement is among the most striking features of quantum mechanics
\cite{sc35} and a prerequisite for quantum computation and simulation
schemes \cite{nich00,ekjo98}. Typically, for dissipation to induce
entanglement, the dissipative processes require multi-quanta exchange
with the environment in addition to carefully engineered coupling
constants. Realizations of artificial system-environment couplings
are, in general, hard to achieve. On the other hand, it has been shown
that multi-quanta dissipation is a natural component in nonlinear
systems \cite{dykr84}. For instance, graphene-based nanomechanical
resonators possess a strong intrinsic nonlinearity \cite{voki+12} and
exhibit nonlinear dissipation \cite{vocr+13}. In other setups the
conservative nonlinearity has to be induced by coupling to auxiliary
systems\cite{ja07sefu+09}. Similarly, nonlinear dissipation can be
induced in opto-mechanical systems \cite{lemi04,nubo+10,taba+13}, and
has also been proposed to emerge in superconducting solid state
quantum devices \cite{evsp+12}.

For a single oscillator mode of a nonlinear system under the influence
of two-quanta dissipation, resulting from a nonlinear coupling to the
reservoir, it was shown that nonclassical steady states with non-zero
coherences can be generated \cite{silo78,gikn93,giga+94,lo84,vocr+13}.
Moreover, the formation of superposition states \cite{evsp+12,taba+13}
as well as non-thermal and squeezed states were shown to be obtainable
by two-phonon cooling \cite{nubo+10}.

Here, we investigate two coupled nonlinear oscillators, which are
individually subject to two-quantum dissipation. We start from a
microscopic description of the total system and derive a quantum
master equation (QME) for the reduced density matrix. By using the
rotating wave approximation (RWA) we show that in the weak coupling
limit, the two-quantum dissipation mainly influences the short-time
dynamics. Additionally, dephasing is affecting the system in the
long-time limit. By numerically solving the QME and the QME in the
RWA, we demonstrate that at zero temperature entanglement between the
oscillators is created, which persists even if the coupling is
switched off. For sufficiently large temperatures, we find early stage
disentanglement (ESD), also known as sudden death of
entanglement\cite{yueb09}.

\section{Model} 
We consider two nonlinear oscillators, which are linearly coupled. 
Additionally, each oscillator is quadratically coupled to a reservoir
consisting of harmonic oscillators. Accordingly, the total
Hamiltonian is $H = H_{\rm S}+ H_{\rm B}+ H_{\rm SB}$, where
$\hbar=1$, the masses of the oscillators are $m=1$, and
\begin{subequations}\label{eq:hamiltonian_terms}
\begin{align}
H_{\rm S} = & \sum_{m=1,2}
\left(\frac{1}{2}p_m^2+\frac{1}{2}\omega_m^2 q_m^2 +
\frac{2\mu_m}{3}q_m^4 \right)\nonumber\\ 
 & + \sqrt{\omega_1 \omega_2}\lambda q_1 q_2
\;,\label{eq:hamiltonian_terms2}\\ H_{\rm B} = &\sum_m\sum_k \omega_{m
  k} b^{\dag}_{m k} b_{m k}\;,\label{eq:hamiltonian_terms3}\\ H_{\rm
  SB} = &\sum_m q_m^2 \sum_k (2 \omega_m \eta_{m k}) (b^{\dag}_{m k} + b_{m
  k})\;.\label{eq:hamiltonian_terms4}
\end{align}
\end{subequations}
Here, $p_m=\imath \sqrt{\omega_m/2}(a_m^{\dag}-a_m)$ and
$q_m=(a_m^{\dag}+a_m)/\sqrt{(2 \omega_m)}$ denote momentum and
oscillation amplitude of oscillator $m$, respectively, and
$a_m^{(\dag)}$ is the anihilation (creation) operator of the $m$'th
oscillator. The oscillators are characterized by their frequencies
$\omega_m$, the strength of the nonlinearities $\mu_m$ and the
coupling strength $\lambda>0$. The operator $b^\dagger_{mk}$
($b_{mk}$) creates (destroys) a phonon in state $k$ of reservoir $m$
with the frequency $\omega_{m k}$. The coupling strength of oscillator
$m$ to reservoir state $k$ is denoted by $\eta_{m k}$. 
The nonlinear coupling in (\ref{eq:hamiltonian_terms4}) 
will lead to nonlinear damping in the 
classical limit \cite{dykr84}. For nanomechanical systems such a coupling is realized, 
for example, by coupling of in-plane and flexural motion \cite{crmi+12}; 
in opto-mechanical setups \cite{nubo+10,taba+13,lemi04} and in superconducting solid 
state quantum devices \cite{evsp+12} it can be induced by using an auxiliary system.

In the weak system-reservoir coupling limit, the evolution of the
reduced density matrix $\RDM$ is given by a QME. Following the
standard approach by using the Born-Markov approximation in the
interaction picture with respect to $H_{\rm S}$ one obtains
\cite{brpe02}
\begin{eqnarray}\label{eq:QMEwIntfinal}
\frac{\partial }{\partial t}\rho(t) =& 
-\sum\limits_{l,m}\int\limits_0^{\infty} d\tau\Big[ 
S_l(t) , S_m(t-\tau)\rho(t)
\Big]
C_{l m}(\tau) \nonumber\\
&\:-
\Big[\:
S_l(t),\rho(t) S_m(t-\tau)
\Big]
C_{m l}(-\tau).
\end{eqnarray}
Here, the operators $S_m(t) = e^{\imath H_{\rm S} t}\;(a_m^{\dag} +
a_m)^2 \;e^{-\imath H_{\rm S} t}$ and $B_m(t)= \sum_k{\eta_{m
    k}}\left(b^{\dag}_{m k} e^{\imath \omega_{m k} t} + b_{m k}
e^{-\imath \omega_{m k} t}\right)$ were used to decompose the coupling
Hamiltonian $H_{\rm SB}(t)= \sum_{m=1,2} S_m(t)\otimes B_m(t)$ in the
interaction picture. Assuming the reservoirs to be initially in
thermal equilibrium, $\rho_{\rm B}=\rho_{{\rm B},1}\otimes\rho_{{\rm
    B},2}$, the reservoir correlation functions $C_{m l}(\tau) =
\Tr_{\rm B} \{B_m(t) B_l(t-\tau)\rho_{\rm B} \}$ are given by
\begin{multline}\label{eq:corrfunct}
	C_{m l}(\tau) = \delta_{m l}
        \int\frac{d\omega}{2\pi}\kappa_{m}(\omega) \left[ N(\omega)
          e^{\imath \omega \tau} \right. \\ \left.+ (N(\omega) + 1)
          e^{-\imath \omega \tau} \right]\;,
\end{multline}
where $N(\omega)=(e^{\omega/k_{\rm B}T}-1)^{-1}$ is the Bose-Einstein
distribution and $\kappa_{m}(\omega)=2\pi \sum_k |\eta_{m
  k}|^2\delta(\omega-\omega_{m k})$ is the spectral density. The
specific form of $\kappa_m$ depends on the microscopic details of the
system-reservoir coupling. If $\kappa_m$ is sufficiently smooth around
the frequencies of interest, the exact frequency dependence is not
crucial. To be specific, we use an Ohmic spectral density,
$\kappa_{m}(\omega)=\Gamma_m \omega/(2\omega_m)$, where $\Gamma_m$ is
the nonlinear dissipation strength.

In many systems the time-scales associated with $\omega_m$ and $\mu_m$
are well separated. In this case the RWA can be used to simplify the
QME. For convenience we consider a symmetric setup in the following,
i.e., $\omega_1=\omega_2=\omega_0$ and $\Gamma_1=\Gamma_2=\Gamma_0$.

Further, we define the one-sided
Fourier transform of the reservoir correlation function
\begin{equation}\label{eq:gamma_def}
	\frac{1}{2} \gamma_m (\omega)
        +\imath \sigma_m (\omega)\; = \int\limits_0^{\infty} d\tau\; e^{\imath
          \omega \tau} C_{mm}(\tau).
\end{equation}
The rates $\gamma_m$ determine the strength of dissipation, while $\sigma_m$
renormalizes the system Hamiltonian. Using the
expression of the bath correlation function \eqref{eq:corrfunct} one
finds that
\begin{subequations}
\begin{align}
	\gamma_m(2\omega_0) ={}&  \Gamma_0 [N(2\omega_0) + 1]\;,\\
	\gamma_m(-2\omega_0) ={}& \Gamma_0 N(2\omega_0)\;.
\end{align}
\end{subequations}

In the weak coupling limit, $\lambda \ll \omega_0$, one finally
obtains the QME in RWA
\begin{align}\label{eq:RWA_QME}
	\frac{\partial }{\partial t}\rho(t) ={}& -\imath \left[ H_{\rm
            RWA}, \rho \right] + \sum_m \left( \gamma_m(2\omega_0)
        \mathcal{L}[a^\dag_m a^\dag_m] \right.\notag\\ {}&\left.  +
        \gamma_m(-2\omega_0) \mathcal{L}[a_m a_m] \right) \rho +
        \mathcal{D}_{12}(\lambda) \rho
\end{align}
in Schr\"{o}dinger representation. The oscillator Hamiltonian in RWA is given by
\begin{align}\label{eq:RWAHam}
	H_{\rm RWA} ={}& \sum_m \left( (\omega_0+\mu_m) a^\dag_m a_m +
        \mu_m a^\dag_m a_m a^\dag_m a_m \right) \notag\\ {}&+
        \frac{\lambda}{2} \left( a^\dag_1 a_2 + a^\dag_2 a_1 \right)
\end{align}
and the superoperator $\mathcal{L}$ is defined as
\begin{equation}
	\mathcal{L}[X]\rho = -\frac{1}{2} X X^\dag \rho -
        \frac{1}{2} \rho X X^\dag + X^\dag \rho X\;.
\end{equation}
Thus, to lowest order in $\lambda$, the oscillators are individually
coupled to their respective reservoirs. The other superoperator
$\mathcal{D}_{12}(\lambda)$ becomes
\begin{align}
	\mathcal{D}_{12}(\lambda)\rho
	={}& \Upsilon_{+} \mathcal{L}[(n_1 - n_2)]\rho
        \notag\\ {}& -\frac{1}{2}\Upsilon_{-} \Big[ (n_1 - n_2)
          (a^\dag_1 a_2 - a^\dag_2 a_1) \rho \Big.\notag\\ {}&-
          (a^\dag_1 a_2 - a^\dag_2 a_1) \rho (n_1 - n_2)
          \notag\\ {}&+ \rho (a^\dag_1 a_2 - a^\dag_2 a_1)^\dag
          (n_1 - n_2)\notag\\ {}&\Big.- (n_1 - n_2) \rho (a^\dag_1
          a_2 - a^\dag_2 a_1)^\dag \Big]\;, \label{eq:cross_D}
\end{align}
where $\Upsilon_{\pm} = \gamma(\lambda)\pm\gamma(-\lambda)$ with
$\gamma(\lambda) = \kappa_0(\lambda) [N(\lambda) + 1]$ and
$\gamma(-\lambda) = \kappa_0(\lambda) N(\lambda)$.

The terms in the QME \eqref{eq:RWA_QME} which are proportional to
$\gamma_m(2\omega_0)$ describe the loss of two quanta into the bath,
while the terms which are proportional to $\gamma_m(-2\omega_0)$ give
rise to the absorption of two quanta from the bath. For zero
temperature only the former processes are present, since $N(2\omega_0)
= 0$. In contrast to the contributions involving
$\gamma_m(\pm2\omega_0)$, the superoperator $\mathcal{D}_{12}$
contains two dephasing terms, proportional to $\Upsilon_+$ and
$\Upsilon_-$, respectively. One sees that these two terms have a
different temperature dependence. For $\lambda\ll k_{\rm B} T$,
$\Upsilon_{+} > \Upsilon_{-}$, while for $\lambda\gg k_{\rm B} T$ one
has $\Upsilon_{+} \approx \Upsilon_{-}$. Note that, if the
oscillators were linearly coupled to the reservoirs, the latter
dephasing terms would only arise if $\Gamma_1 \neq \Gamma_2$
\cite{pool+05}.

\section{Results} 
\subsection{Zero Temperature}
\begin{figure*}[ht]
  \centering
  \includegraphics[width=0.72\textwidth]
             {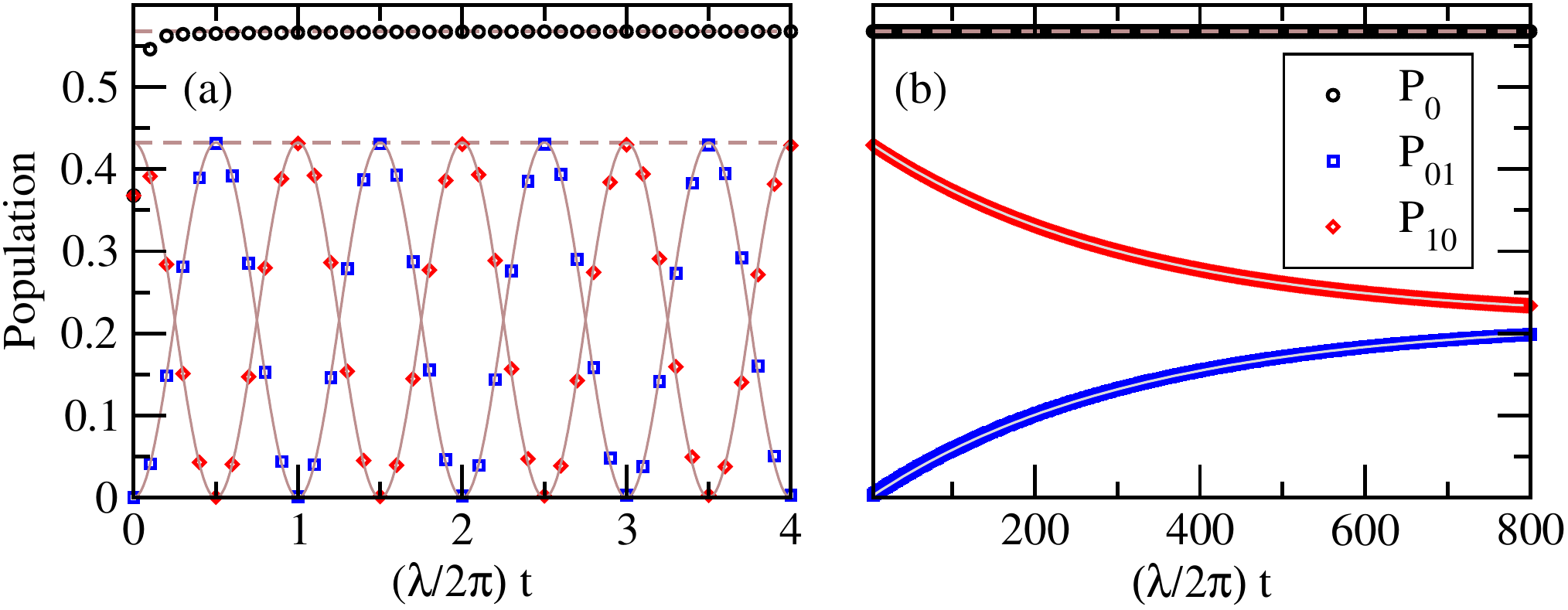}\hfill
  \includegraphics[width=0.28\textwidth]
             {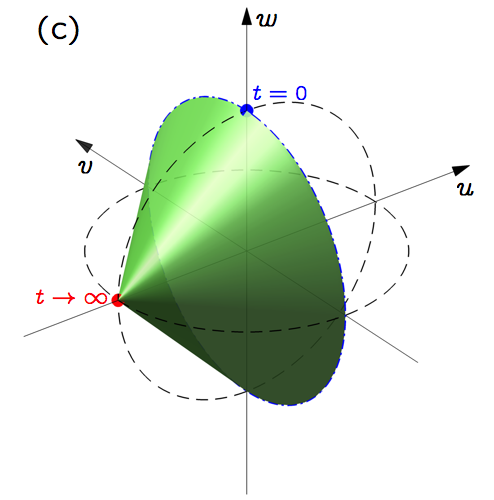}            
\caption{(Color online) Time dependence of the populations of the
  states $\ket{0,0}$, $\ket{1,0}$ and $\ket{0,1}$. The initial
  oscillator amplitudes are $\alpha_1=1$ and $\alpha_2 =0$. For the
  calculation $M=10$ states are taken into account for each
  oscillator. The nonlinear dissipation strength is
  $\Gamma_0=10^{-3}$, the coupling strength between the oscillators is
  $\lambda=\Gamma_0/2$, and $\mu=\Gamma_0$. Thin lines in (a) and (b)
  are given by Eqs.\ \eqref{eq:anasol}. For visual convenience only
  the envelope is shown in (b). (c) Dynamics of the Bloch vector
  $(u,v,w)$ according to Eqs.\ \eqref{eq:anasol}. The cone indicates
  the states which are attained during the time-evolution.  }
  \label{fig:zeroT_Population}
\end{figure*}
First we concentrate on $T=0$, which implies
$\gamma_m(-2\omega_0)=\gamma(-\lambda)=0$. We introduce the basis vector
$\ket{n,i} = \ket{n}_1 \otimes \ket{i}_2$, which denotes a state with
$n$ quanta in oscillator $1$ and $i$ quanta in oscillator $2$. From
Eq.\ \eqref{eq:RWA_QME} one sees that a density matrix, which involves
the states $\ket{0,0}$, $\ket{0,1}$ and $\ket{1,0}$ will not be
affected by the dissipation and the QME does not lead to a coupling to
other states. In other words, the steady-state will, in general, not
be the ground state $\ket{0,0}$, but rather a mix of superpositions of
$\ket{0,0}$, $\ket{0,1}$ and $\ket{1,0}$. This state can be written as
\begin{multline}\label{eq:stat_state}
	\RDM = P_{00} \ket{0,0}\bra{0,0} + P_{10}\ket{1,0}\bra{1,0} +
        P_{01} \ket{0,1}\bra{0,1} \\ + \Big( \RDM_{00,10} \ket{0,0}\bra{1,0}
        + \RDM_{00,01} \ket{0,0}\bra{0,1} \\ +\RDM_{10,01}
        \ket{1,0}\bra{0,1} + {\rm h.c.}\Big),
\end{multline}
where weights of the respective states are determined by the initial
state and evolve according to the Hamiltonian $H_{\rm RWA}$ and the
superoperator $\mathcal{D}_{12}$. For example, the populations $P_{00}$
and $P_{10}+P_{01}$ can be obtained from the sum of the populations of
the even ($P_{\rm even}$) and odd ($P_{\rm odd}$) states in the
initial density matrix\footnote{ The definitions are $P_{\rm even} =
  \sum_{n,i; n+i\; \text{even}} \rho_{ni,ni}$ and $P_{\rm odd} =
  \sum_{n,i; n+i\; \text{odd}} \rho_{ni,ni}$.  }, respectively. This
is a result of the invariance of the total Hamiltonian to changes of
total parity ($q_1 \to -q_1$ and $q_2 \to -q_2$). A similar behavior
is observed for a single oscillator \cite{silo78,vocr+13}, where it
leads to the formation of a nonclassical state with non-vanishing
coherences.

To investigate the dynamics of the coupled oscillators the QME
\eqref{eq:QMEwIntfinal} and the QME in RWA \eqref{eq:RWA_QME} are solved
numerically. The Hilbert space is truncated after $M=10$
states for each oscillator. Equation \eqref{eq:QMEwIntfinal} is solved
in the eigenbasis of the system Hamiltonian, which corresponds to the
Wangsness-Bloch-Redfield method \cite{wabl53bl57re65}. Initially,
the system is prepared in a product state of two coherent states,
$\psi(0)=\ket{\alpha_1}\otimes\ket{\alpha_2}$, where $\ket{\alpha_m} =
\exp(\alpha_m a_m^\dagger - \alpha_m^* a_m)\ket{0}_m$, $\alpha_m =
\sqrt{2}q_m(0)$ is the amplitude of initial displacement, and the rest of
the system parameters are $\omega_0=1$,
$\mu_m=\mu=\Gamma_0=10^{-3}$ and $\lambda=\Gamma_0/2$.

The dynamics of the populations of the states $\ket{0,0}$, $\ket{1,0}$
and $\ket{0,1}$ is shown in Fig.\ \ref{fig:zeroT_Population}(a) for
the situation where only one oscillator is initially displaced
($\alpha_1=1$ and $\alpha_2 =0$). After a transient behavior up to
$t\approx2\pi/4\lambda$, the population of the ground state settles to
a constant value $P_{00}=P_{\rm even}$ and oscillations of $P_{10}$
and $P_{01}$ can be seen, while $P_{01}+P_{10}=P_{\rm odd}$ remains
constant. In the long-time limit, $P_{01}=P_{10}$ due to the presence
of dephasing. This is shown in Fig.\ \ref{fig:zeroT_Population}(b).
The transient behavior corresponds to the individual relaxation of
each oscillator from the initial state to a state which has at most
one excitation per oscillator.

Using the QME in RWA \eqref{eq:RWA_QME} the
equations of motion for the matrix-elements in the steady-state
\eqref{eq:stat_state} can be solved. It is convenient to introduce
$s(t)=P_{01}+P_{10}$ and the components of the Bloch vector
$w(t)=P_{10}-P_{01}$, $u(t)=\RDM_{01,10} + \RDM_{10,01}$ and
$v(t)=-\imath(\RDM_{10,01} - \RDM_{01,10})$. In the limit
$\lambda\gg\Upsilon_{+}=\Upsilon_{-}\equiv\Upsilon$, their solutions
are
\begin{subequations}\label{eq:anasol}
\begin{align}
	s(t) =& P_{\rm odd}, \\
	w(t) =& e^{-\Upsilon t}[w_0 \cos(\lambda t)-v_0\sin(\lambda t)], \\
	u(t) =& P_{\rm odd}(e^{-2\Upsilon t} -1 )+ u_0 e^{-2\Upsilon t}, \\
	v(t) =& e^{-\Upsilon t}[v_0 \cos(\lambda t)+w_0\sin(\lambda t)] \;.
\end{align}
\end{subequations}
In the long time limit, $w\to0$ and $v\to0$ while $u\to -P_{\rm odd}$,
due to dephasing. During its time-evolution, the Bloch vector traces
the surface of a cone as can be seen in
Fig.\ \ref{fig:zeroT_Population}(c). Without dephasing ($\Upsilon=0$)
it describes a circle.

\begin{figure}[b]
  \centering
      \includegraphics[width=0.48\textwidth]
                      {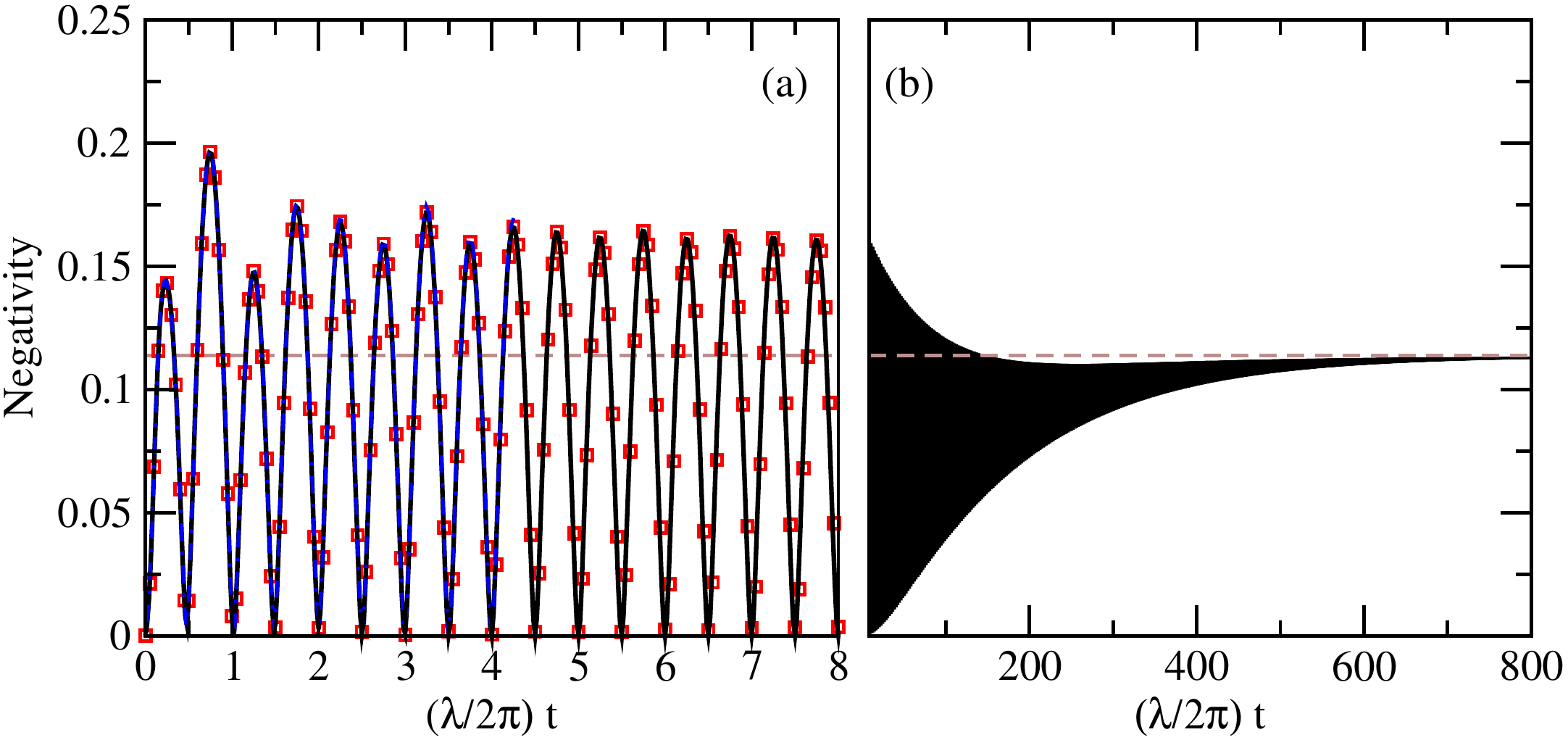}\\                      
      \includegraphics[width=0.48\textwidth]
                      {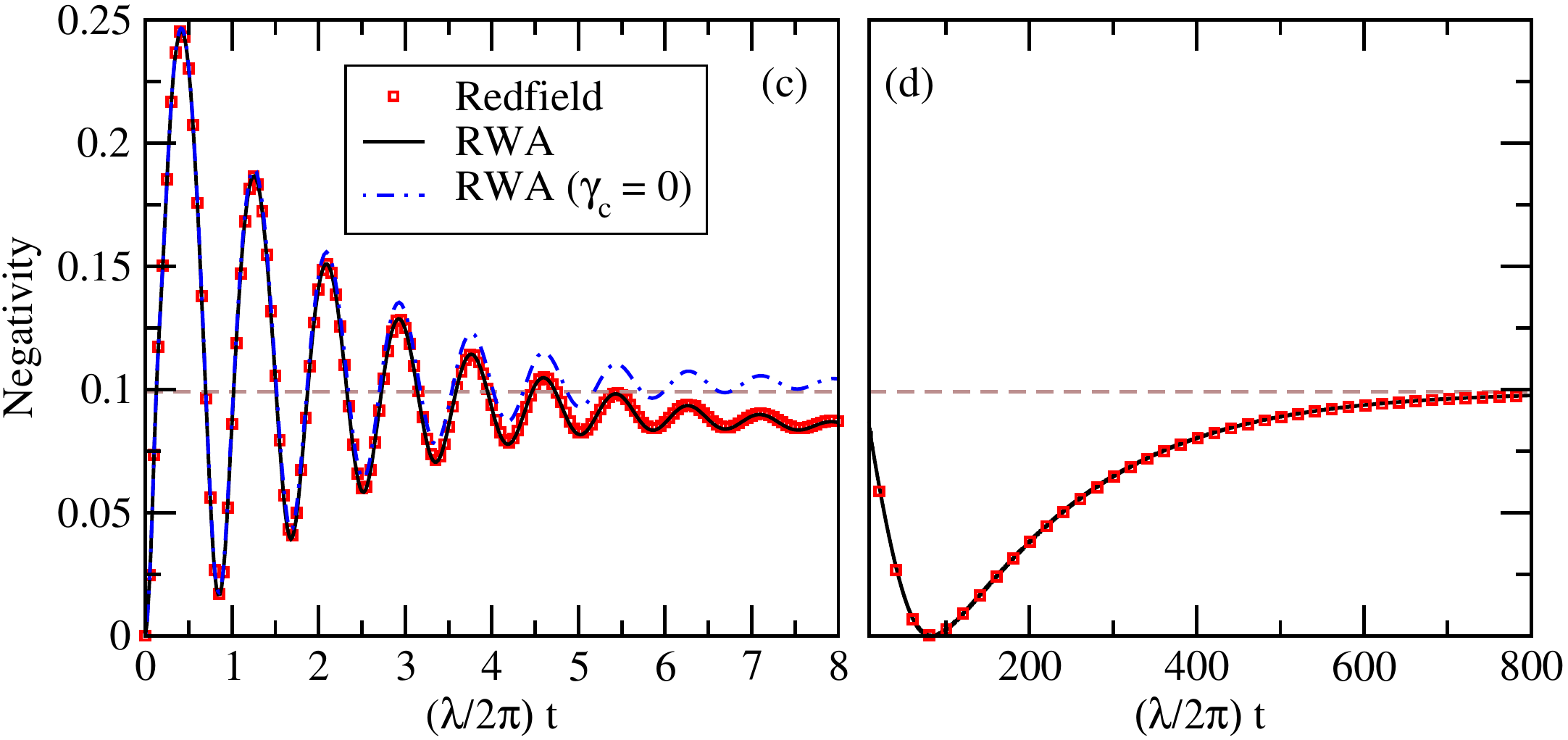}                
\caption{(Color online) Time dependence of the negativity. The initial
  oscillator amplitudes are (a,b) $\alpha_1=1$ and $\alpha_2 =0$ and
  (c,d) $\alpha_1=1$ and $\alpha_2 =1$. The rest of the parameters are
  chosen as in Fig. \ref{fig:zeroT_Population}. Symbols and full lines
  denote numerical results from the Redfield approach and the RWA,
  respectively. The dashed lines indicate the asymptotic
  negativity.\label{fig:zeroT_Negativity}}
\end{figure}
To quantify the entanglement of the oscillators, which is created by
coupling the two systems, the negativity $\mathcal{N} =
({\norm{\RDM^{T_{1}}}-1})/2$ is computed for each time-step
\cite{viwe02}. The matrix $\RDM^{T_{1}}$ is the partial transpose of
the bipartite mixed state $\RDM$ with respect to oscillator $1$. The
negativity corresponds to the absolute value of the sum of negative
eigenvalues of $\RDM^{T_{1}}$ and vanishes for any separable
state. 

For the steady state $\eqref{eq:stat_state}$ the negativity is
found from the negative roots of the characteristic polynomial of the
state's partially transposed density matrix.
According to Vieta's formula, the product of the four roots is
\begin{equation}\label{eq:4roots}
z_1 z_2 z_3 z_4 = -\vert \rho_{01,10} \vert^2 P_{10} P_{01}. 
\end{equation}
This implies that at least one root $z_i$ has to be negative if
$P_{10}$, $P_{01}$ and $\rho_{01,10}$ are non-vanishing, which results
in a finite negativity. 

The numerically obtained behavior of
$\mathcal{N}$ for the case with $\alpha_1=1$ and $\alpha_2 =0$ is
shown in Figs.\ \ref{fig:zeroT_Negativity}(a) and (b). An almost
periodic behavior is observed, with a vanishing negativity at
multiples of a half-period $2\pi/2\lambda$. Comparing with
Fig.\ \ref{fig:zeroT_Population}(a) one finds that these times
correspond to maximal population in either $\ket{1,0}$ or $\ket{0,1}$,
which happens twice per period. At those times the coherence between
$\ket{1,0}$ and $\ket{0,1}$ vanishes. On the other hand, maximal
negativity is found when the coherence is maximal and $P_{10}=P_{01}$,
which also happens twice per period. It is important to realize that
the entanglement persists even if the the coupling is switched off,
since the steady state \eqref{eq:stat_state} is not affected by the
nonlinear dissipation, and the dephasing depends on the presence of
the coupling $\Upsilon \propto \lambda$.

The time-evolution of the negativity is mainly governed by the
dynamics of $\vert \rho_{01,10}\vert^2 = \vert u/2 \vert^2$. The fast
oscillations in Fig. \ref{fig:zeroT_Negativity}(b) can be understood
by the circular revolution of the Bloch vector around the $u$-axis in
Fig. \ref{fig:zeroT_Population}(c). The negativity dip can in the same
manner be understood in terms of $u(t)$ initially being in the
vicinity of the phase space origin in addition to the requirement of
Eq.\ \eqref{eq:4roots} being fulfilled. The negativity saturation is
due to $u(t)$ reaching its final value at the tip of the cone as $t\to
\infty$.

In the case of two displaced oscillators, $\alpha_1=\alpha_2 =1$, the
populations $P_{00},P_{01}$ and $P_{10}$ again undergo a short
transient behavior and quickly saturate at finite, steady values. The
time evolution of the negativity is shown in
Figs.\ \ref{fig:zeroT_Negativity}(c) and (d). Initially, the
negativity has decaying oscillations, with a period
$\approx2\pi/\lambda$, which occur due to the influence of decaying
coherences between states with more than one excitation. For finite
$\Upsilon$, the negativity first decays and then increases to settle
at a finite value. This behavior is governed by $\Re\rho_{01,10} =
u/2$, which is initially positive and monotonically decreases
according to Eq.\ \eqref{eq:anasol}. Since $w=v=0$ the Bloch vector
only points along the $u$-axis. When it crosses the origin, the
negativity is minimal. This is also the reason for the absence of
long-term oscillations in Fig.\ \ref{fig:zeroT_Negativity}(d). The
asymptotic value of the negativity is given by
\begin{equation}\label{eq:neg_2disp}
	\mathcal{N}_{\infty} = \frac{1}{2}\left( P_{\rm odd} - 1 + \sqrt{(
          P_{\rm odd} - 1)^2 + P_{\rm odd}^2} \right)\;,
\end{equation}
which only depends on the initial state via $P_{\rm odd}$. If
$\Upsilon$ were zero, $u(t)=u(0)$, and the negativity would quickly
saturate at a finite value, as seen in
Fig.\ \ref{fig:zeroT_Negativity}(c).

\begin{figure}[b]
  \centering
      \includegraphics[width=0.48\textwidth]
                      {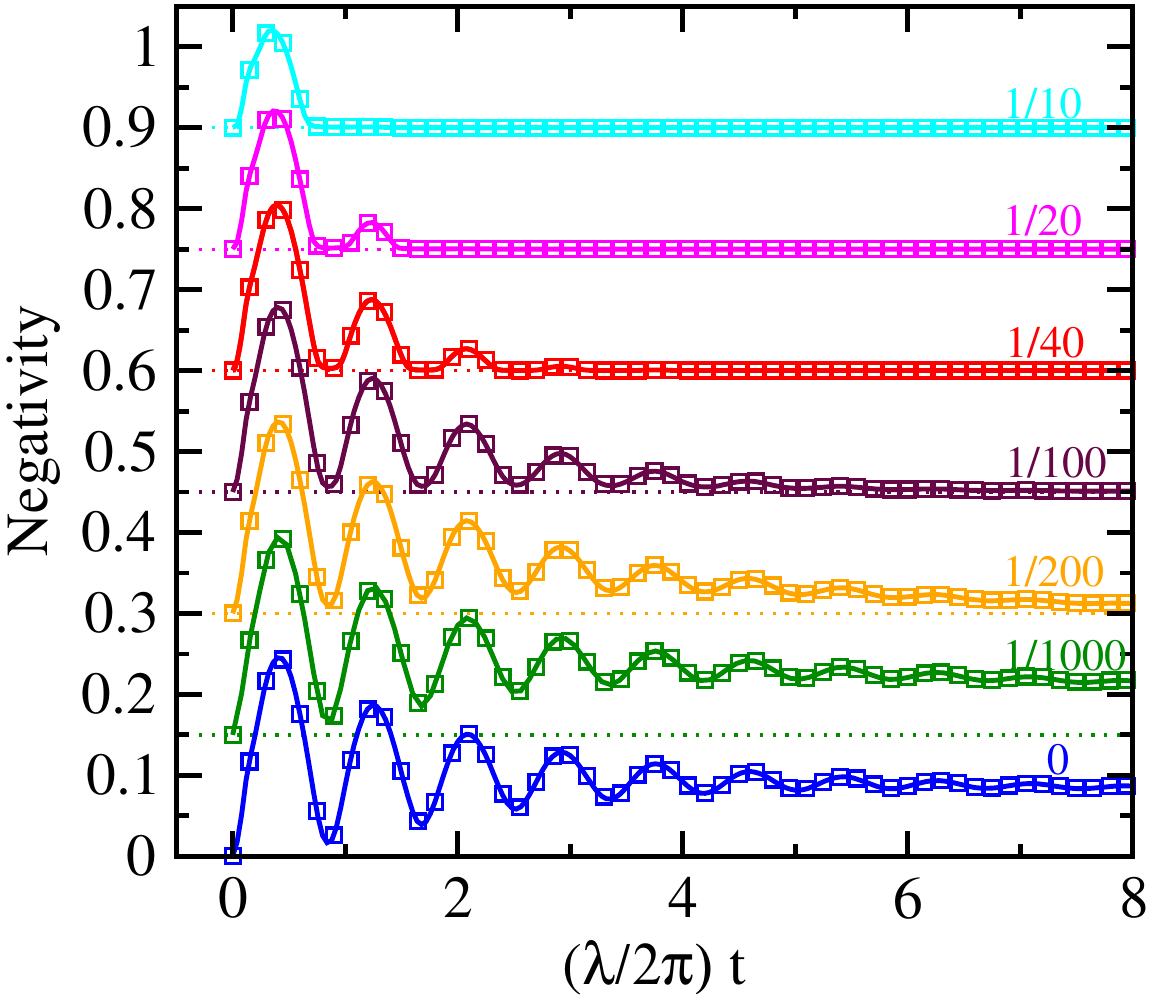}
\caption{(Color online) Time and temperature dependence of the
  negativity. The initial oscillator amplitudes are $\alpha_1=\alpha_2
  =1$. The rest of the parameters are chosen as in
  Fig. \ref{fig:zeroT_Population}. The curves are shifted by $0.15$
  for convenience. The temperatures in units of $\Omega/k_{\rm B}$ are
  given by the labels next to the curves.
  \label{fig:finT_Negativity}}
\end{figure}
\subsection{Finite Temperature} 
At finite temperatures the thermal excitations created by the bath
will, in general, lead to a decay of the coherences between the Fock
states. Therefore, one expects to observe a decay of the entanglement
with increasing temperature. In Fig.\ \ref{fig:finT_Negativity} the
time and temperature dependence of the negativity is shown. For
temperatures $k_{\rm B}T/\omega_0<1/100$ the negativity is slowly
decaying with time. For larger temperatures the negativity is seen to
be zero after a finite time. This behavior is known as ESD
\cite{yueb09}. These results are also consistent with the previous
result \cite{alja08}, that the occurrence of ESD and of entanglement
sudden (re)birth at finite temperatures are generic features.

\section{Summary} 
In this work we have investigated the quantum dynamics of two
weakly-interacting anharmonic oscillators, which are nonlinearly
coupled to individual dissipative environments. This scenario leads to
the formation of non-classical steady states and, in particular,
entanglement. Additionally, at finite temperatures the exotic features
of entanglement sudden death and (re)birth are observed. Our results
show that dissipation-induced quantum state generation is feasible
without engineering the system-environment coupling. Utilizing the
natural presence of nonlinear dissipation in nonlinear nano-scale
systems provides a promising route to realize exotic quantum state
generation. 

\begin{acknowledgements}
The research leading to this article has received funding from the
Swedish Research Council (VR).
\end{acknowledgements}


\end{document}